\documentclass[12pt]{article}
\usepackage{amsmath}
\usepackage{amssymb}
\usepackage{theorem}
\usepackage{color}
\usepackage{epic}
\newcommand{\R}{{\cal R}}
\newcommand{\B}{{\cal B}}
\newcommand{\F}{{\cal F}}
\newcommand{\A}{{\cal A}}
\newcommand{\Nb}{{\mathbb{N}}}
\newcommand{\Qb}{{\mathbb{Q}}}
\newcommand{\Rb}{{\mathbb{R}}}

\newcommand{\xmb}{{\mathbf{x}}}
\newcommand{\ymb}{{\mathbf{y}}}
\newcommand{\zmb}{{\mathbf{z}}}

\newcommand{\pmbold}{{\mathbf{p}}}
\newcommand{\lambdamb}{{\mbox{\boldmath\(\lambda\)}}}
\newcommand{\qed}{\hfill\hbox{\rule{6pt}{6pt}}}
\newtheorem{theorem}{Theorem}[section]
\newtheorem{lemma}{Lemma}[section]
\newtheorem{col}{Corollary}[section]
\newtheorem{prop}{Proposition}[section]

{\theorembodyfont{\rmfamily}\newtheorem{remark}{Remark}
{\theorembodyfont{\rmfamily}\newtheorem{example}{Example}
\DeclareMathOperator*{\argmax}{arg\,max}

\newcommand{\edit}[1]{\textcolor{black}{#1}}

\begin{document}
\title{Algorithmic randomness and monotone complexity on product space}
\author{Hayato Takahashi\\
The Institute of Statistical Mathematics\\
10-3 Midori-cho, Tachikawa, Tokyo 190-8562, Japan\\
hayato.takahashi@ieee.org}
\date{\today}
\maketitle
\begin{abstract}
We study algorithmic randomness and monotone complexity on product of the set of infinite binary sequences. 
We explore the following problems: monotone complexity on product space, Lambalgen's theorem for correlated probability,
classification of random sets by likelihood ratio tests, decomposition of complexity and independence, Bayesian statistics for individual random sequences. 
Formerly Lambalgen's theorem for correlated probability is shown under a uniform computability assumption in [H.~Takahashi Inform.~Comp.~2008].
In this paper we show the theorem without the assumption. 
\end{abstract}

\section{Introduction}
It is known that Martin-L\"of random sequences \cite{martin-lof66}  satisfy many laws of probability one, for example ergodic theorem, martingale convergence theorem, and so on, see \cite{{vyugin98},{takahashiIandC}}.
In this paper, we study Martin-L\"of random sequences with respect to a probability on product space \(\Omega\times\Omega\), where \(\Omega\) is the set of infinite binary sequences. 
In particular, we investigate the following problems:
\begin{enumerate}
\item Randomness and monotone complexity on product space (Levin-Schnorr theorem for product space)

\item Lambalgen's theorem \edit{\cite{lambalgen87}}  for correlated probability. 

\item  Likelihood ratio test and classification of random sets. 

\item  Decomposition of complexity and independence of individual random sequences.

\item Bayesian statistics for individual random sequences.

\end{enumerate}
The above problems are property of product space except for 3.

In Section~\ref{sec-lambalgen}, we show Lambalgen's  theorem for correlated probability.
In the previous paper \cite{takahashiIandC}, the theorem is shown under a uniform computability assumption.
In this paper, we show the theorem without that assumption. 
This is the main theorem of this paper (Theorem~\ref{th-general-Lambalgen}).

The other sections are as follows:
In Section~\ref{sec-random}, we define monotone complexity on product space. 
A usual definition of one-dimensional monotone complexity strongly depends on an order structure of one-dimensional space.
\edit{In order to define monotone complexity on product space,} we give an algebraic definition of monotone \edit{function} for product space, which is applicable, {\it mutatis mutandis}, to an abstract partially ordered set.
In Section~\ref{sec-ratio-test}, we show a classification of random sets by likelihood ratio tests.
In particular we show an important theorem by  Martin-L\"of, i.e.,  two computable probabilities are mutually singular iff their random sets are disjoint.
As a simple application, we show consistency of MDL for individual sequences. 
In Section~\ref{sec-decomp-complexity}, 
we show a decomposition of monotone complexity for prefixes of random sequences under a condition. 
As a corollary, we show some equivalent conditions for \edit{independence of}  individual random sequences.
In Section~\ref{sec-bayes}, we  apply our results to Bayesian statistics. 
By virtue of randomness theory, we can develop a point-wise theory for Bayesian statistics. 
In particular, we show consistency of posterior distribution (and its equivalent conditions) for individual random sequences. 
In order to show this, the results of Section~\ref{sec-ratio-test} plays an important role. 
Also we show an asymptotic theory of estimation for individual sequences, which is closely related to decomposition of complexity. 

\section{Randomness and complexity}\label{sec-random}
First we introduce Martin-L\"of randomness on \(\Omega\).
Let \(S\) be the set of finite binary strings. Let \(\Omega\) be the set of infinite binary sequences with product topology.
As in \cite{takahashiIandC}, we write \(A\subset B\) including \(A=B\). 
\edit{Throughout the paper, the base of logarithm is 2. }
We use symbols such as \(x,y,s\) to denote an element of \(S\) and  \(x^\infty,y^\infty\) to denote an element of \(\Omega\).
For \(x\in S\), let \(\Delta(x):=\{x\omega : \omega\in\Omega\}\), where \(x\omega\) is the concatenation of \(x\) and \(\omega\), and for \(x^\infty\in\Omega\), 
\(\Delta(x^\infty):=\{x^\infty\}\).
Let \(\lambda\in S\) be the empty word, then \(\Delta(\lambda)=\Omega\).
\edit{For \(A\subset S\),  let \(\sigma\{\Delta(x)\}_{x\in A}\) be the \(\sigma\)-algebra generated by \(\{\Delta(x)\}_{x\in A}\) and \(\B:=\sigma\{\Delta(x)\}_{x\in S}\)}.
Let \((\Omega,\B,P)\) be a probability space. 
We write \(P(x):=P(\Delta(x))\) for \(x\in S\), then we have \(P(x)=P(x0)+P(x1)\) for all \(x\).
Let \(\Nb\), \(\Qb\), and \(\Rb\) be the set of natural numbers, rational numbers, and real numbers, respectively. 
\(P\) is called computable if there exists a computable function \(p: S\times\Nb\to\Qb\) such that \(\forall x\in S\forall k\in\Nb\ |P(x)-p(x,k)|<1/k\).
A set \(A\subset S\) is called recursively enumerable (r.e.) if there is a computable function \(f: \Nb\to S\) such that \(f(\Nb)=A\).
For \(A\subset S\), let \(\tilde{A}:=\cup_{x\in A}\Delta(x)\).
A set \(U\subset \Nb\times S\) is called (Martin-L\"of) test with respect to \(P\) if 1) \(U\) is r.e., 2) \(\tilde{U}_{n+1}\subset \tilde{U}_n\) for all \(n\), where \(U_n=\{x : (n,x)\in U\}\), and
3) \(P(\tilde{U}_n)<2^{-n}\). In the following, if \(P\) is obvious from the context, we say that \(U\) is a test. 
A test \(U\) is called universal if for any other test \(V\), there is a constant \(c\) such that \(\forall n\ \tilde{V}_{n+c}\subset \tilde{U}_n\).
\begin{theorem}[Martin-L\"of\cite{martin-lof66}]\label{th-martin}
If \(P\) is a computable probability, a universal test \(U\) exists.
\end{theorem}
In \cite{martin-lof66}, the set \((\cap_{n=1}^\infty \tilde{U}_n)^c\) (complement of the limit of universal test) is defined to be random sequences with respect to \(P\), where \(U\) is a universal test. 
We write \(\R^P:=(\cap_{n=1}^\infty \tilde{U}_n)^c\).
Note that  for two universal tests \(U\) and \(V\), \(\cap_{n=1}^\infty \tilde{U}_n=\cap_{n=1}^\infty \tilde{V}_n\) and hence \(\R^P\) does not depend on the choice of a universal test.
An equivalent definition of test is that  \(U\) is r.e. and \(\sum_n P(\tilde{U}_n)<\infty\). Then the set  covered by \(\tilde{U}_n\) infinitely many times is a limit of a test, i.e.,
\(\limsup_n \tilde{U}_n\subset (\R^P)^c\), see \cite{shen89}.

For \(x,y\in S\), let \(\Delta(x,y):=\Delta(x)\times\Delta(y)\).
Let \(\B_{S^2}:=\sigma\{\Delta(x,y)| x,y\in S\}\).
Then computability of \(P\) on \((\Omega^2,\B_{S^2})\), its Martin-L\"of tests, and the set of random sequences are defined similarly.

\subsection{Complexity}\label{sub-sec-complexity}
For \(x',x\in S\cup\Omega\), we write \(x'\sqsubseteq x\Leftrightarrow x'\) is a prefix of \(x\Leftrightarrow \Delta(x')\supset\Delta(x)\),
and for \((x',y'), (x,y)\in (S\cup\Omega)^2\), \((x',y')\sqsubseteq (x,y)\Leftrightarrow x'\sqsubseteq x\) and \(y'\sqsubseteq y\) \(\Leftrightarrow \Delta(x',y')\supset\Delta(x,y)\).
Then \(S\cup\Omega\) and \((S\cup\Omega)^2\) are partially ordered sets. 
For \(A\subset S^2\), let \(\bigvee A\) be the least upper bound of \(A\).
Then \(\bigvee A\) exists in \((S\cup\Omega)^2\) iff \(\cap_{(x,y)\in A}\Delta(x,y)\ne\emptyset\).
In the following bold-faced symbols \(\xmb,\ymb,\pmbold\) denote an element of \((S\cup\Omega)^2\), \(\xmb^\infty\) denote an element of \(\Omega^2\), and \(\lambdamb=(\lambda,\lambda)\).

First we define  monotone functions \((S\cup\Omega)^2\to(S\cup\Omega)^2\).\\
Let \(F\subset S^2\times S^2\) and \(F_{\pmbold}:=\{\xmb | (\pmbold, \xmb)\in F\}\).\\
Assume that 
\begin{equation}\label{monotone-condition}
\forall \pmbold\in S^2,\  \lambdamb\in F_\pmbold\mbox{ and } \bigvee_{\pmbold'\sqsubseteq\pmbold} F_{\pmbold'}\mbox{ exists.}
\end{equation}
Set
\begin{equation}\label{def-by-F}
f (\pmbold):= \bigvee_{\pmbold'\sqsubseteq\pmbold,\  \pmbold'\in S^2}
F_{\pmbold'}\mbox{ for } \pmbold\in  (S\cup\Omega)^2.
\end{equation}
We see that  \(f:(S\cup\Omega)^2 \to (S\cup\Omega)^2\) and \(f\) is monotone, i.e., 
\[\pmbold'\sqsubseteq \pmbold\Rightarrow f(\pmbold')\sqsubseteq f(\pmbold).\]
Conversely, let \(f:(S\cup\Omega)^2\to (S\cup\Omega)^2\) be a monotone function, and set
\[F:=\{ (\pmbold,\xmb)\in S^2\times S^2 | \xmb\sqsubseteq f(\pmbold)\},\]
Then \(\bigvee F_{\pmbold} =f(\pmbold)\),  \(F\) satisfies (\ref{monotone-condition}), and  the function defined by \(F\) coincides with \(f\). 
If  \(F\) is a r.e.~set, the function \(f\)  defined by (\ref{def-by-F}) is called {\it computable monotone function}. 

For \(s\in S\), let \(|s|\) be the length of \(s\). In particular  \(|\lambda|=0\) and \(|x^\infty|=\infty\). 
For \(\pmbold=(p_1,p_2)\in (S\cup\Omega)^2\), let \(|\pmbold|:=|p_1|+|p_2|\).
The monotone complexity with respect to a computable monotone function \(f:(S\cup\Omega)^2\to (S\cup\Omega)^2\) is defined as follows:
\[Km^2_f(x,y):=\min\{ |p_1|+|p_2|~|~(x,y)\sqsubseteq f(p_1,p_2)\},\]
\[Km_f(x,y):=\min\{ |p|~|~(x,y)\sqsubseteq f(p,\lambda)\},\]
for \(x,y,p,p_1,p_2\in S\cup\Omega\).
If there is no \((p_1,p_2)\) such that \((x,y)\sqsubseteq f(p_1,p_2)\), then \(Km^2_f(x,y):=\infty\). 
Similarly, \(Km_f(x,y):=\infty\) if there is no \(p\) such that \((x,y)\sqsubseteq f(p,\lambda)\).

A computable monotone function \(u:(S\cup\Omega)^2\to(S\cup\Omega)^2\) is called {\it optimal} 
if for any computable monotone function \(f:(S\cup\Omega)^2\to(S\cup\Omega)^2\), there is a constant \(c\) such that 
\(Km^2_u(\xmb)\leq Km^2_f(\xmb)+c\) for all \(\xmb\in (S\cup\Omega)^2\).
We can construct an optimal function in the following manner.
First, observe that there is a r.e.~set \(\bar{F}\subset \Nb\times  S^2\times S^2\) such that 1)  \(F_i=\{ (\pmbold, \xmb)| (i, \pmbold,\xmb)\in \bar{F}\}\)  satisfies (\ref{monotone-condition}) for all 
\(i\in\Nb\), and 2) for each r.e.~set \(F\) that satisfies (\ref{monotone-condition}),
there is \(i\) such that \(F=F_i\). Note that the first condition in (\ref{monotone-condition}) is necessary to enumerate \(\{F_i\}\).
Next, set \(F^u:=\{(\bar{i}\pmbold,\xmb) | (i,\pmbold, \xmb)\in \bar{F}\}\), where \(\bar{i}\pmbold=(0^i1p_1,p_2)\) for \(\pmbold=(p_1,p_2)\).
Let \(u\) be  a computable monotone function defined by \(F^u\) via (\ref{def-by-F}), then  we see that \(u\) is optimal.
In the following discussion, we fix \(u\) and let
\[Km^2(x,y):=Km^2_u(x,y),\ Km(x,y):=Km_u(x,y),\]
\[Km(x|y):=\min\{ |p|~|~(x,\lambda)\sqsubseteq u(p,y)\},\]
\[Km(x):=Km(x|\lambda)\mbox{ for }x,y\in S\cup\Omega.\]
By definition, we have \(\forall x,y,\ Km^2(x,y)\leq Km(x,y)\).  
Note that \(Km\) is equivalent to a monotone complexity that is defined from an optimal monotone function \(S\cup\Omega\to (S\cup\Omega)^2\).
Also note that \(Km(x)\) defined above is different from  \(Km^2(x):=Km^2(x,\lambda)\).
Later we show that \(Km^2\) and \(Km\) are asymptotically bounded for prefixes of random sequences under a condition, see Corollary~\ref{col-coding}.

\edit{In the following,  a subset  \(\A\) of \(S\cup\Omega\) or \((S\cup\Omega)^2\) is called {\it non-overlapping}  if \(\Delta(\xmb)\cap\Delta(\ymb)=\emptyset\)  for \(\xmb,\ymb\in\A, \xmb\ne\ymb\). 
Note that  \(\Delta(\xmb)\cap\Delta(\ymb)=\emptyset\Rightarrow\) \(\xmb\) and \(\ymb\) are incomparable. 
The converse is true if  \(\xmb,\ymb\in S\cup\Omega\). 
However if \(\xmb,\ymb\in (S\cup\Omega)^2\) then there is a counter-example, e.g.,  \((\lambda, 0)\) and \((0,\lambda)\) are incomparable but \(\Delta(\lambda, 0)\cap\Delta(0,\lambda)=\Delta(0,0)\).
In one-dimensional case,  the notion of non-overlapping is equivalent to that of prefix-free. 
Throughout the paper we use the term ``non-overlapping".
}

\begin{prop}\label{prop-basic-property-km}
a) monotonicity:
\(x\sqsubseteq z\Rightarrow Km(x | y)\leq Km(z|y)\), and \(y\sqsubseteq z\Rightarrow Km(x|y)\geq Km(x|z)\).\\
b) Kraft inequality: \( \sum_{\xmb\in \A} 2^{-Km(\xmb)}\leq\sum_{\xmb\in \A} 2^{-Km^2(\xmb)}\leq 1\) for \edit{non-overlapping} set \(\A\subset (S\cup\Omega)^2\).\\
c) Conditional sub-additivity: \(\exists c~\forall x,y\in S\cup\Omega,\ Km^2(x,y)\leq Km(x|y)+Km(y)+c\).
\end{prop}
Proof)
a) Obvious. 
b) 
Let \(u\) be an optimal monotone function and \(p_{\xmb}\in\{\pmbold | \xmb\sqsubseteq u(\pmbold)\}\).
Suppose that 
 \(\Delta(\xmb)\cap\Delta(\xmb')=\emptyset\) and \(\exists \zmb, \zmb=p_{\xmb}\vee p_{\xmb'}\).
Then \(\xmb\sqsubseteq u(\zmb)\) and \(\xmb'\sqsubseteq u(\zmb)\), which contradicts to  \(\Delta(\xmb)\cap\Delta(\xmb')=\emptyset\).
Thus \(\{p_{\xmb} | \xmb\in\A\}\) is \edit{non-overlapping} for a \edit{non-overlapping} set \(\A\).
By setting \(p_{\xmb}\) to be an optimal code, i.e., \(|p_{\xmb}|=Km^2_u(\xmb)\),  we have \( \sum_{\xmb\in \A} 2^{-Km^2(\xmb)}\leq 1\).
Since \(Km^2\leq Km\), we have the statement.
c) 
Let \(u\) be an optimal monotone function.
Suppose that \(x\sqsubseteq u(p,y)\), \(Km(x|y)=|p|\) and \(y\sqsubseteq u(p'), Km(y)=|p'|\).
Let  \(f:(S\cup\Omega)^2\to(S\cup\Omega)^2\)  such that \(f(p_1,p_2):= (u(p_1,u(p_2)),u(p_2))\) for all \(p_1,p_2\).
Then  \(f\) is monotone and \(Km^2_f(x,y)\leq |p|+|p'|=Km(x|y)+Km(y)\).
\qed

Next we show  Levin-Schnorr theorem  for product space.
Let  \(\A\subset S^2\) be a r.e.~set and 
\edit{\[\A(\xmb^\infty):=\{\xmb\in\A \mid \xmb\sqsubset\xmb^\infty\}\mbox{ for }\xmb^\infty\in\Omega^2.\]}
Before proving the theorem, we need  conditions on \(\A\):
\begin{equation}\label{cond-partition-1}
\xmb,\ymb\in \A\Rightarrow \xmb\mbox{ and }\ymb\mbox{ are comparable or }\Delta(\xmb)\cap\Delta(\ymb)=\emptyset.
\end{equation}
If (\ref{cond-partition-1}) holds then for any \(\A'\subset\A\) there is a \edit{non-overlapping} \(\A''\subset\A'\) such that  \(\tilde{\A}''=\tilde{\A}'\).
\edit{
Note that it is possible \(\A''\) is not r.e.~even if \(\A'\) is a r.e.~set.
\begin{equation}\label{cond-partition-2}
\begin{aligned}
\xmb,\ymb\in \A\Rightarrow\exists\text{ non-overlapping }\alpha\subset\A,\ \Delta(\xmb)\cap(\Delta(\ymb))^c=\tilde{\alpha}.
\end{aligned}
\end{equation}
\begin{lemma}\label{L-S-lem}
If \(\A\) is r.e.~and satisfies  (\ref{cond-partition-2}) then for any r.e.~\(\A'\subset\A\) there is a non-overlapping r.e.~\(\A''\subset\A\) such that \(\tilde{A}'=\tilde{A}''\).
\end{lemma}
Proof)
Since \(\A'\) is r.e., there is a computable \(a':\Nb\to\A'\) such that \(a'(\Nb)=\A'\).
Let \(\A''(0)=\emptyset\). Suppose that \(\A''(n-1)\) is a finite non-overlapping subset of \(\A\) and \(\tilde{A}''(n-1)=\cup_{1\leq i\leq n-1}\Delta(a'(n))\).
Since \(\A''(n-1)\) is finite, from (\ref{cond-partition-2}), there is a non-overlapping \(\alpha(n)\) such that 
\begin{equation}\label{L-S-lem:A}
\tilde{\alpha}(n)=\Delta(a'(n))\cap (\tilde{\A}''(n-1))^c.
\end{equation}
Since \(\Delta(a'(n))\cap (\tilde{\A}''(n-1))^c\) is compact and \(\alpha(n)\) is non-overlapping, from Heine-Borel Theorem, we see that \(\alpha(n)\) is finite. 
Let 
\(\beta(n):=\{ \zmb\in\A\mid \Delta(\zmb)\subset\Delta(a'(n))\cap (\tilde{\A}''(n-1))^c\}\).
Since \(\A\) is r.e.~and \(\A''(n-1)\) is finite, \(\beta(n)\) is r.e.~from \(a'(n)\) and \(\A''(n-1)\).
In particular, since \(\alpha(n)\subset\beta(n)\), we can compute a finite non-overlapping \(\alpha(n)\) that satisfies  (\ref{L-S-lem:A}) from \(a'(n)\) and \(\A''(n-1)\).
Let \(\A''(n):=\A''(n-1)\cup\alpha(n)\) then \(\A''(n)\) is a finite non-overlapping set. Let \(\A'':=\cup_n \A''(n)\).
By induction, \(\A''\subset\A\) is a non-overlapping r.e.~set such that \(\tilde{A}'=\tilde{A}''\).
\qed
}
\begin{theorem}[Levin-Schnorr theorem \cite{{levin73},{schnorr73},{schnorr77}} on product space]\label{th-levin-schnorr-multi}
Let \(P\) be a computable probability on \((\Omega^2,\B_{S^2})\).
Let  \(\A\) be a r.e.~set that satisfies  \edit{(\ref{cond-partition-1}) and (\ref{cond-partition-2}).} Then
\begin{align*}
\xmb^\infty\in\R^P & \Leftarrow \sup_{\xmb\in \A(\xmb^\infty)} -\log P(\xmb)-Km(\xmb)<\infty,\ \xmb^\infty=\vee\A(\xmb^\infty).\\
\xmb^\infty\in\R^P & \Rightarrow  \sup_{\xmb\in \A(\xmb^\infty)} -\log P(\xmb)-Km(\xmb)<\infty.
\end{align*}
The above statements hold for \(Km^2\).
\end{theorem}
Proof)
Suppose that  \(\xmb^\infty\notin\R^P\) and \(\xmb^\infty=\vee\A(\xmb^\infty)\).
\edit{
Then there is a test \(U\) such that for all \(n\), \(\xmb^\infty\in\tilde{U}_n\) and \(P(\tilde{U}_n)<2^{-n}\).
Let \(U'_n:=\{\ymb\in\A | \exists\xmb\in U_n, \xmb\sqsubseteq\ymb\}\). 
Since \(U_n\) and \(\A\) are r.e.~sets, \(U'_n\subset\A\) is a r.e.~set.
From Lemma~\ref{L-S-lem}, there is a non-overlapping r.e.~set \(U''_n\subset\A\) such that \(\tilde{U}'_n=\tilde{U}''_n\).
Since \(\xmb^\infty=\vee\A(\xmb^\infty)\), we have \(\xmb^\infty\in\tilde{U}''_n\) and \(\forall n, U''_n\cap\A(\xmb^\infty)\ne\emptyset\).
}
Let \(P'\) be a measure such that \(P'(\xmb)=P(\xmb)2^n\) for \(\xmb\in U''_n\) and 0 otherwise.
Since \(P(\tilde{U}''_n)<2^{-n}\), we have \(\sum_{\xmb\in U''_n}P'(\xmb)<1\).
By applying Shannon-Fano-Elias coding to \(P'\) on \(U''_n\), we have
\(\exists c_1,c_2>0\forall n\exists \xmb\in\A(\xmb^\infty)\ Km(\xmb)\leq -\log P(\xmb)-n+K(n)+c_1\leq -\log P(\xmb)-n+2\log n +c_2\),
where \(K\) is the prefix complexity. 

Conversely, let 
\(U_n:=\{\edit{\xmb\in\A\mid} Km(\xmb)< -\log P(\xmb)-n\}\).
From (\ref{cond-partition-1}),
we see that there is a \edit{non-overlapping} set \(U'_n\subset U_n\) such that \(\tilde{U}'_n=\tilde{U}_n\).
Hence
\(P(\tilde{U}_n)=P(\tilde{U}'_n)< \sum_{\xmb\in U'_n}2^{-Km(\xmb)-n}\leq 2^{-n}\), where the last inequality follows from Proposition~\ref{prop-basic-property-km} b.
Since \(U_n\) is a r.e.~set, \(\{U_n\}\) is a test and \(\cap_n \tilde{U}_n\subset (\R^P)^c\).
The proof for \(Km^2\) is the same as above. 
\qed
\begin{example}
Let \(g:\Nb\to\Nb\) be a total-computable monotonically increasing function\edit{, where \(n\leq m\Rightarrow g(n)\leq g(m)\).}
Let 
\begin{equation}\label{def:Ag}
\A_{\edit{g}}:=\{(x,y)\in S^2 \mid |y|=g(|x|)\}.
\end{equation}
Then \(\A_{\edit{g}}\) is decidable and \edit{satisfies (\ref{cond-partition-1}) and (\ref{cond-partition-2}).}
If \(g\) is unbounded then \(\forall \xmb^\infty,  \vee\A_{\edit{g}}(\xmb^\infty)=\xmb^\infty\).
\end{example}
Next we study a coding problem for multi-dimensional monotone complexity. 
The following lemma shows that if \(\A\) is decidable and satisfies (\ref{cond-partition-1}), we have the same  one-dimensional coding as in \cite{USS90}.
\begin{lemma}\label{lem-coding}
Let \(P\) be a computable probability on \((\Omega^2,\B_{S^2})\) and 
let \(\A\subset S^2\) be a decidable set that satisfies (\ref{cond-partition-1}), then
there is a computable monotone function \(g: S\cup\Omega\to (S\cup\Omega)^2\) such that 
\[\exists c \forall \xmb\in A(\xmb^\infty), Km_g(\xmb)\leq -\log P(\xmb)+c.\]
\end{lemma}
Proof)
If \(\A\) is decidable and satisfies (\ref{cond-partition-1}) then,  
by rearranging an enumeration of \(\A\), we see that 
there is a computable \(f:\Nb\to S^2\) such that \(f(\Nb)=\A\) and 
\(\forall i,j,\ i<j,\ \Delta(f(i))\cap\Delta(f(j))=\emptyset \mbox{ or }f(i)\sqsubseteq f(j)\).
Then we can construct a family of half-open intervals \(V_{f(i)}:=[a(i),b(i))\subset [0,1], \forall  i\in \Nb\) that satisfies the following conditions:
0) \(V_\lambda=[0,1]\),
1) \(| V_{f(i)}|=P(f(i))\) for all \(i\), where \(|V|\) is the length of the interval \(V\), 
2) if \(\Delta(f(i))\cap\Delta(f(j))=\emptyset\) then \(V_{f(i)}\cap V_{f(j)}=\emptyset\),
3) if \(f(i)\sqsubseteq f(j)\) then \(V_{f(i)}\supset V_{f(j)}\),
and 4) \(a\) and \(b\) are computable, i.e., there are rational valued computable functions \(A:\Nb\times\Nb\to\Qb\) and \(B:\Nb\times\Nb\to\Qb\)
such that \(\forall i,k,\ |a(i)-A(i,k)|<1/k,\ |b(i)-B(i,k)|<1/k\).
For \(s=s_1s_2\cdots s_n\in S, \forall i, s_i\in\{0,1\}\),
let \(I_s:=[\sum_{1\leq i\leq n} s_i2^{-i}, \sum_{1\leq i\leq n} s_i2^{-i}+2^{-n})\).
Then set
\(F:=\{(s,f(i))\in S\times S^2 | I_s\subset V_{f(i)}, i\in\Nb\}\).
We see that \(F\) is a r.e.~set that satisfies (\ref{monotone-condition}).
Let \(g\) be a computable monotone function defined by \(F\), then we have
\(g: S\cup\Omega\to (S\cup\Omega)^2\) and 
\(\exists c \forall \xmb\in A(\xmb^\infty), Km_g(\xmb)\leq -\log P(\xmb)+c\).
\qed

From Theorem~\ref{th-levin-schnorr-multi}, Lemma~\ref{lem-coding}, and  Proposition~\ref{prop-basic-property-km} c, we have
\begin{col}\label{col-coding}
Let \(P\) be a computable probability on \((\Omega^2,\B_{S^2})\).
If \(\A\subset S^2\) is decidable and satisfies \edit{(\ref{cond-partition-1}) and (\ref{cond-partition-2})}, then 
\begin{align*}
\xmb^\infty\in\R^P & \Rightarrow\sup_{\xmb\in \A(\xmb^\infty)}  |\log P(\xmb)+Km(\xmb)|<\infty,\\
\xmb^\infty\in\R^P & \Leftarrow\sup_{\xmb\in \A(\xmb^\infty)}  |\log P(\xmb)+Km(\xmb)|<\infty, \xmb^\infty=\vee\A(\xmb^\infty).
\end{align*}
The above statements are true for \(Km^2\), and 
\begin{align*}
\xmb^\infty\in\R^P & \Rightarrow \sup_{\xmb\in \A(\xmb^\infty)}  |Km(\xmb)-Km^2(\xmb)|<\infty\\
& \Rightarrow \sup_{(x,y)\in \A(\xmb^\infty)} Km(x,y)-Km(x|y)-Km(y)<\infty.\nonumber
\end{align*}
\end{col}

For 1-dimensional monotone complexity and its relation to other complexities, see \cite{LV97,UspenskyShen96}.
In \cite{cho5}, a conditional complexity \(K_\ast\) that is monotone with  the conditional argument is defined.
\begin{remark}\label{rem-abstract}
It is not difficult to develop monotone function and complexity in an abstract way.
Indeed, let \(A\) and \(\bar{A}\) be partially ordered sets such that \(A\) is r.e.~and \(\bar{A}:=\{\vee B | B\subset A\}\).
Let \(F\subset A\times A\) be a r.e.~set that satisfies (\ref{monotone-condition}) with respect to the partial  order of \(A\).
Then we can define (optimal) monotone function \(f:\bar{A}\to\bar{A}\) \edit{in a similar way with Section~\ref{sub-sec-complexity}}. 
For example, for  \(\xmb,\ymb\in(S\cup\Omega)^\infty\), let \(\xmb\sqsubseteq\ymb\) if \(\forall i, x^i\sqsubseteq y^i\) for \(\xmb=(x^1,x^2,\ldots), \ymb=(y^1,y^2,\ldots), x^i,y^i\in S\cup\Omega\).
Then  \((S\cup\Omega)^\infty\) is a partially ordered set. 
\edit{
Let \(A:=\{ (\xmb,\lambda^\infty) | \xmb\in\cup_k S^k\}\), where \(\lambda^\infty=(\lambda,\lambda,\ldots)\in S^\infty\).
Then \(A\) is a sub-partially ordered set of \((S\cup\Omega)^\infty\)  and  \(\bar{A}=(S\cup\Omega)^\infty\). We can define computable monotone function \(f:\bar{A}\to\bar{A}\).
For \(\xmb=(x_1,\ldots,x_n,\ldots)\in \bar{A}\), let \(|\xmb|:=\sum_n |x_n|\).
Then  \(Km_f\) is defined.}
\edit{For example, let us consider discrete time (computable) stochastic processes \(X_i\in\Omega, i=1,2,\ldots\).
Then their randomness and complexity of sample paths are modeled with a computable probability on  \((\Omega^\infty,\B_A)\) and \(Km_f\), where \(\B_A:=\sigma\{\Delta(\xmb)| \xmb\in A\}, \Delta(\xmb):=\{\xmb^\infty | \xmb\sqsubset\xmb^\infty\in\Omega^\infty\}\) and 
computability of  probabilities on \((\Omega^\infty,\B_A)\) is defined in a similar manner with finite dimensional case.}
\end{remark}
\edit{
\begin{remark}\label{rem-abstract2}
Let \(\phi^{l,t}:(S\cup\Omega)^l\to (S\cup\Omega)^t\) be an optimal monotone function for \(1\leq l,t\leq\infty\).
Then \(Km^{l,t}(x_1,\ldots, x_t)\leq Km^{l',t}(x_1,\ldots, x_t)+O(1)\) if \(l'\leq l\), where \(Km^{l,t}\) is defined from \(\phi^{l,t}\).
If  \(\A\subset S^t, t<\infty\) or \(\A\subset \{ (\xmb,\lambda^\infty) | \xmb\in\cup_k S^k\}, t=\infty\) is a decidable set that satisfies  (\ref{cond-partition-1}) and (\ref{cond-partition-2}) then
Theorem~\ref{th-levin-schnorr-multi} and Corollary~\ref{col-coding} hold for \(Km^{l,t}\) for \(1\leq l,t\leq\infty\). 
In order to simplify the argument, in the following discussion, we use \(Km\).
\end{remark}
}

\section{Section and relativized randomness}\label{sec-lambalgen}

Let \(P\) be a computable probability on \(X\times Y=\Omega^2\).
Let \(P_X\) and \(P_Y\) be its marginal distributions on \(X\) and \(Y\), respectively, i.e.,
\(P_X(x)=P(x,\lambda)\) and \(P_Y(y)=P(\lambda,y)\) for \(x,y\in S\).
Let 
\[P(x\vert y):=\left\{
\begin{array}{rl}
\frac{P(x,y)}{P_Y(y)},& \ \mbox{if }P_Y(y)>0\\
0,& \ \mbox{if }P_Y(y)=0
\end{array}\right.,\]
and
\[P(x\vert y^\infty):=\lim_{y\to y^\infty}P(x\vert y),\]
for \(y^\infty\in\Omega\) if the right-hand side exists.
For a subset \(A\subset X\times Y\) and \(y^\infty\in Y\), set
\[A_{y^\infty}:=\{x^\infty\vert (x^\infty,y^\infty)\in A\}.\]
For example, \(\R^P_{y^\infty}=\{x^\infty | (x^\infty,y^\infty)\in\R^P\}\).
Similarly, for  \(B\subset S\times S\), set \(B_{y^\infty}:=\{x | (x,y)\in B, y\sqsubset y^\infty\}\).
\begin{theorem}[\cite{takahashiIandC}]\label{th-ex-cond}
If \(y^\infty\in\R^{P_Y}\), then 
\(P(x\vert y^\infty)\) exists for all \(x\in S\), and 
\(P(\cdot\vert y^\infty)\) is a probability measure on   \((\Omega,\B)\).
\end{theorem}
\begin{theorem}[\cite{takahashiIandC}]\label{col-sectionA}
 \(P(\R^P_{y^\infty}\vert y^\infty)=1\) if \(y^\infty\in\R^{P_Y}\). \(\R^P_{y^\infty}=\emptyset\) if \(y^\infty\notin\R^{P_Y}\).
 \end{theorem}
 \begin{col}[\cite{takahashiIandC}]\label{col-sectionB}
 \(\R^{P_X}=\cup_{y^\infty\in\R^{P_Y}}\R^P_{y^\infty}\).
 \end{col}
 
If  \(P(\cdot\vert y^\infty)\) is computable relative to \(y^\infty\), then let \(\R^{P(\cdot |y^\infty),y^\infty}\) be the set of random sequences with respect to \(P(\cdot\vert y^\infty)\) relative to \(y^\infty\).
\edit{In \cite{takahashiIandC}, \(\{P(\cdot | y^\infty)\}_{y^\infty}\) is called uniformly computable if  
there is a partial computable \(A\) such that \(\forall y^\infty\in\R^{P_Y}, x\in S, k\in\Nb\exists y\sqsubset y^\infty,\ |P(x|y^\infty)-A(x,y,k)|<1/k\), i.e., \(P(\cdot\vert y^\infty)\) is uniformly computable relative to all  \(y^\infty\in\R^{P_Y}\).}
In \cite{takahashiIandC}, it is shown that \(\R^{P(\cdot |y^\infty),y^\infty}\subset\R^P_{y^\infty}\), and under uniform computability, \(\R^{P(\cdot |y^\infty),y^\infty}=\R^P_{y^\infty}\) 
for \(y^\infty\in\R^{P_Y}\). 
In the following we show the equivalence without assuming the uniform computability\edit{; we only assume that \(P(\cdot |y^\infty)\) is computable relative to  a given \(y^\infty\in\R^{P_Y}\).
In order to show \(\R^{P(\cdot |y^\infty),y^\infty}\supset\R^P_{y^\infty}\), first we extend a test \(U^{y^\infty}_n\) w.r.t.~\(P(\cdot |y^\infty)\)   to a test w.r.t.~a finite measure \(P'\) on \(\Omega^2\) such that 
the section of the extended test at \(y^\infty\) coincide with \(U^{y^\infty}_n\) and the total measure of the extended test w.r.t.~\(P'\) is sufficiently small.
Finally by using Markov inequality, we construct a test w.r.t.~\(P\).}

\begin{theorem}\label{th-general-Lambalgen}
Assume that \(y^\infty\in\R^{P_Y}\) and \(P(\cdot\vert y^\infty)\) is computable relative to \(y^\infty\), then \(\R^{P(\cdot |y^\infty),y^\infty}=\R^P_{y^\infty}\).
\end{theorem}
Proof)
Fix \(y^\infty\in\R^P_Y\). 
Since \(P(\cdot\vert y^\infty)\) is computable relative to \(y^\infty\), there is a partial computable function \(A: S\times S\times \Nb\to \{q\in\Qb | q\geq 0\}\) such that 
\edit{(a1)} 
\(
\forall x,k \exists y\sqsubset y^\infty, |P(x|y^\infty)-A(x,y,k)|<\frac{1}{k}
\)
\edit{and (a2)} if \(A(x,y,k)\) is defined then \(A(x,y,k)=A(x,z,k)\) for all \(y\sqsubseteq z\).
Similarly, let \(U^{y^\infty}\subset \Nb\times S\) be a Martin-L\"of test with respect to \(P(\cdot\vert y^\infty)\) relative to \(y^\infty\), i.e., \(U^{y^\infty}\) is a r.e.~set relative to \(y^\infty\),  
 and \(P(\tilde{U}^{y^\infty}_n | y^\infty)<2^{-n}\) for all \(n\), where \(U^{y^\infty}_n:=\{x | (n,x)\in U^{y^\infty}\}\).
Then there is a partial computable function \(B: \Nb\times\Nb\times S\to S\) such that \edit{(b1)} \(\forall n,\ U^{y^\infty}_n=\{x | \exists i,y\sqsubset y^\infty, B(i,n,y)=x\}\) \edit{and (b2)}
if \(B(i,n,y)\) is defined then \(B(i,n,y)=B(i,n,z)\) for all \(y\sqsubseteq z\).

Let \(U_n:=\{(x,y) | \exists i,\ B(i,n,y)=x\}\). 
Then \(U_{n,y^\infty}=U^{y^\infty}_n\).
Let  \(U'_n\subset S\times S\) be a \edit{non-overlapping} r.e.~set
such that \(\tilde{U}_n=\tilde{U}'_n\).
Then \(\tilde{U}'_{n,y^\infty}=\tilde{U}^{y^\infty}_n\).
Let 
\begin{align}\label{def-Vn}
V_n:=\{ (x,z,k) \mid &\  (x,y)\in U'_n, \edit{y\sqsubseteq z\in S}, k\in\Nb,\nonumber\\
& \frac{1}{k}<\frac{1}{2} A(x,z,k)\text{ or } (\edit{k\geq 2^{n+|x|}},\ A(x,z,k)<\frac{1}{k})\},
\end{align}
\(V^{X\times Y}_n:=\{(x,y) \mid (x,y,k)\in V_n\}\).
Then we have
\begin{equation}\label{CX:1}
(x,y)\in V^{X\times Y}_n\Rightarrow \forall y\sqsubseteq z, (x,z)\in V^{X\times Y}_n,
\end{equation}
\begin{equation}\label{CX:2}
\forall z^\infty\in\Omega,\ V_{n,z^\infty}^{ X\times Y}\text{ is non-overlapping},
\end{equation}
\begin{equation}\label{C:3}
V_{n,y^\infty}^{ X\times Y}=U'_{n,y^\infty},
\end{equation}
\edit{
where (\ref{CX:1}) follows from (a2); (\ref{CX:2}) follows from that \(U'_n\) is non-overlapping; 
(\ref{C:3}) follows from that: from (a1) and (a2),  (i)   if \(P(x|y^\infty)>0\) then  \(\exists y\sqsubset y^\infty, k\forall y\sqsubseteq z, \frac{1}{k}<\frac{1}{2}A(x,z,k)\) and  (ii)
if \(P(x|y^\infty)=0\) then \(\forall k\exists y\sqsubset y^\infty\forall y\sqsubseteq z\) such that \(A(x,z,k)< \frac{1}{k}\).}

Note that 
 if   \(\frac{1}{k}<\frac{1}{2}A(x,y,k)\) and \(y\sqsubset y^\infty\) then    \(|P(x|y^\infty)-A(x,y,k)|<\frac{1}{k}<\frac{1}{2}A(x,y,k)\), i.e.,
\begin{equation}\label{condition-A0}
\frac{1}{2}A(x,y,k)<P(x|y^\infty)<\frac{3}{2}A(x,y,k).
\end{equation}

From \(V_n\), we can construct a r.e.~set \(W_n\subset S\times S\times \Nb\) that satisfies (\ref{condition-A}), (\ref{condition-A1}), (\ref{condition-B}), (\ref{condition-D}), and (\ref{condition-E}) 
 (Lemma~\ref{lemma-construction} below):
\begin{equation}\label{condition-A}
W_n\subset V_n.
\end{equation}
\begin{equation}\label{condition-A1}
W^{X\times Y}_n\text{ is non-overlapping, where } W^{X\times Y}_n:=\{ (x,y) |  (x,y,k)\in W_n\}.
\end{equation}
\begin{equation}\label{condition-B}
 (x,y,k),(x,y,k')\in W_n\Rightarrow k=k'.
 \end{equation}
 \begin{equation}\label{condition-D}
\forall z^\infty\in\Omega,\ \sum_{(x,y,k)\in W_n, y\sqsubset z^\infty} A(x,y,k)<3\cdot 2^{-n}.
\end{equation}
\begin{equation}\label{condition-E}
\tilde{U}^{y^\infty}_n=\tilde{W}^{X\times Y}_{n,y^\infty}.
\end{equation}

Let \(P'(x,z):=A(x,z,k)P_Y(z)\) for \((x,y,k)\in W_n, y\sqsubseteq z\) and \(P'(x,y):=0\) for \((x,y)\) such that 
\(\Delta(x,y)\cap \tilde{W}^{X\times Y}_n=\emptyset\).
Then by (\ref{condition-D}), \(P'(\tilde{W}^{X\times Y}_n)<3\cdot 2^{-n}\).

Finally let  
\begin{align*}
U^{X\times Y}_n:=\{ (x,z)\in S\times S\  |~& (x,y)\in W^{X\times Y}_n, \ y\sqsubseteq z,\\
& P(x,z)<\frac{3}{2}P'(x,z)\mbox{ or }P(x,z)<\edit{2^{-n-|x|}P_Y(z)}\}.
\end{align*}
Since \(W^{X\times Y}_n\) is r.e.~and \(P\) is computable,  we see that \(U^{X\times Y}_n\) is a r.e.~set.
\edit{Since \(W^{X\times Y}_n\) is non-overlapping, we have \(\sum_{(x,y)\edit{\in W^{X\times Y}_n}}2^{-|x|}P_Y(y)\leq 1\) and 
\[P(\tilde{U}^{X\times Y}_n)<\frac{3}{2}P'(\tilde{W}^{X\times Y}_n)+\sum_{(x,y)\edit{\in W^{X\times Y}_n}}2^{-n-|x|}P_Y(y)< \frac{11}{2}\cdot 2^{-n}.\]
From (\ref{condition-A}), we have \((x,y,k)\in W_n\Rightarrow  \frac{1}{k}<\frac{1}{2} A(x,y,k)\text{ or } k\geq 2^{n+|x|},\ A(x,y,k)<\frac{1}{k}\).
Since \(P(x|y)\to P(x|y^\infty)\) as \(y\to y^\infty\) for \(y^\infty\in\R^{P_Y}\) (Theorem~\ref{th-ex-cond}), we have for \((x,y)\in W^{X\times Y}_n, y\sqsubset y^\infty\) 
(i) if  \(\frac{1}{k}<\frac{1}{2}A(x,y,k)\)  then from (\ref{condition-A0}), \(\exists y\sqsubseteq z\sqsubset y^\infty, P(x,z)<\frac{3}{2}P'(x,z)\) and (ii)
if \(k\geq 2^{n+|x|}, A(x,y,k)<\frac{1}{k}\) then \(\exists y\sqsubseteq z\sqsubset y^\infty, P(x,z)<2^{-n-|x|}P_Y(z)\).}
Thus \(\tilde{W}^{X\times Y}_{n, y^\infty}\subset\tilde{U}^{X\times Y}_{n,y^\infty}\). 
Since \(\tilde{U}^{X\times Y}_n\subset\tilde{W}^{X\times Y}_n\), from  (\ref{condition-E}), we have
\[
\tilde{U}^{y^\infty}_n=\tilde{U}^{X\times Y}_{n,y^\infty}.
\]
\edit{Since \(U^{X\times Y}:=\{(n, x, y) | (x,y)\in U^{X\times Y}_n\}\) is r.e.~and \(\sum_n P(\tilde{U}^{X\times Y}_n)<\infty\)}, we have \(\limsup_n \tilde{U}^{X\times Y}_n \subset (\R^P)^c\) and \(\R^P_{y^\infty}\subset \R^{P(\cdot |y^\infty),y^\infty}\).
The converse inclusion is shown in \cite{takahashiIandC}. \qed

\begin{figure}[ht]
\setlength{\unitlength}{0.8mm}
\begin{picture}(80,70)(-20,0)
\put(35,10){\framebox(60,60)}
\put(95,0){\(Y\)}
\put(25,65){\(X\)}
\dashline{1}(56,32)(56,40)
\put(50,32){\framebox(12,8){{\footnotesize \(\gamma_1~\gamma_2\)}}}
\put(50,50){\framebox(6,8){\(\alpha\)}}
\put(72,27){\framebox(10,13){\(\beta\)}}
\thicklines
\put(50,7){\line(0,1){6}}
\put(56,7){\line(0,1){6}}
\put(72,7){\line(0,1){6}}
\put(82,7){\line(0,1){6}}
\end{picture}
\caption{This figure illustrates a construction of \(W_n\).
For example, suppose that \(W(t-1)=\{ (x_1, y_1, k_1), (x_2, y_2, k_2)\}, \alpha=\Delta(x_1,y_1), \text{ and } \beta=\Delta(x_2,y_2)\) for some \(t\) as shown in  the figure. 
 \(W^Y\) is illustrated by the partition on the \(Y\)-axis. 
If  \(v(t)=(x_3,y_3, k_3)\) and \(\gamma=\Delta(x_3,y_3)\) (the rectangle below \(\alpha\)) then \(\gamma\) is divided into \(\gamma_1=\Delta(x_3, y_1)\) and \(\gamma_2=\Delta(x_3,z')\).
If \( A(x_1, y_1, k_1)+A(x_3, y_1,k_3)<3\cdot 2^{-n}\) then \((x_3,y_1,k_3)\in W(t)\), and if \(A(x_3,z',k_3)<3\cdot 2^{-n}\) then \((x_3,z',k_3)\in W(t)\).
}
\label{fig-0}
\end{figure}
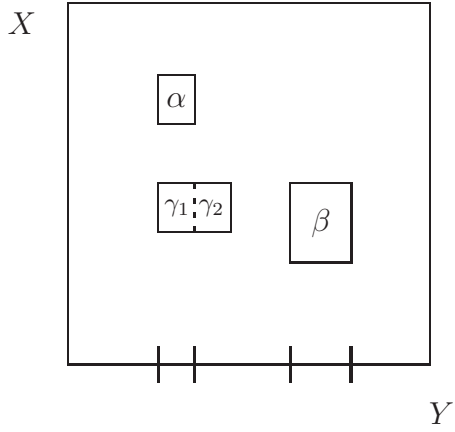

\begin{lemma}\label{lemma-construction}
There is a r.e.~set \(W_n\) that satisfies  (\ref{condition-A}), (\ref{condition-B}), (\ref{condition-D}), and (\ref{condition-E}).
\end{lemma}
Proof)
We construct a r.e.~set \(W_n\subset S\times S\times\Nb\) by induction. 
Let \(W(0):=\emptyset\).
Suppose that \(W(t-1)\subset V_n\) is finite, \(W^{X\times Y}(t-1):=\{(x,z) | (x,z,k)\in W(t-1)\}\) is non-overlapping, and 
\begin{equation}\label{condition-F}
\forall z^\infty\in\Omega,\ \sum_{(x,y,k)\in W(t-1),\ y\sqsubset z^\infty} A(x,y,k)<3\cdot 2^{-n}.
\end{equation}
Since \(W(t-1)\) is finite, there is a finite \edit{non-overlapping} set \(W^Y\) such that \(\cup_{y\in W^Y}\Delta(y)=\Omega\) and \(\sigma\{\Delta(y) | y\in W^Y\}=\sigma\{\Delta(y)| (x,y,k)\in W(t-1)\}\).
Since \(V_n\) is a r.e.~set, let \(v:\Nb\to V_n\) be a computable function such that \(v(\Nb)=V_n\).
Let
\begin{align*}
w(t):=\{(x,z',k)\in S\times S\times\Nb \  |\ &  v(t)=(x,y,k), \\
& \exists z\in W^Y,  z':=y\vee z\mbox{ exists,}\\
& W(t-1)\cup\{(x,z',k)\}\mbox{ satisfies }(\ref{condition-F}),\\
& \edit{ \tilde{W}^{X\times Y}(t-1)\cap\Delta(x,z')=\emptyset}\},
\end{align*}
and \(W(t):=W(t-1)\cup w(t)\).
Let \(w^Y:=\{z | (x,z,k)\in w(t)\}\).
Since \(W^Y\) is non-overlapping, \(w^Y\) is non-overlapping.
\edit{Hence (i) if \((x,z',k)\in w(t)\) and \(z^\infty\in\Delta(z')\) then \(\{(x,y,k) | (x,y,k)\in W(t), y\sqsubset z^\infty\}=\{(x,y,k) | (x,y,k)\in W(t-1), y\sqsubset z^\infty\}\cup\{(x, z',k)\}\) and
(ii) if \(z^\infty\notin \tilde{w}^Y\) then \(\{(x,y,k) | (x,y,k)\in W(t), y\sqsubset z^\infty\}=\{(x,y,k) | (x,y,k)\in W(t-1), y\sqsubset z^\infty\}\), see Figure~\ref{fig-0}.}
Thus  (\ref{condition-F}) holds for \(W(t)\).
By induction, \(W(t)\) is finite and satisfies (\ref{condition-F}) for all \(t\).
Since \(W(t-1)\) is finite, we see that \(w(t)\) is decidable.
Let \(W_n:=\cup_t W(t)\) then \(W_n\) is a r.e.~set.
Since \(\forall t\ W(t-1)\subset W(t)\), from (\ref{condition-F}), we have  (\ref{condition-D}).
From (\ref{CX:1}) we have  (\ref{condition-A}).
From  the last condition of the definition of \(w(t)\), we have  (\ref{condition-A1}) and (\ref{condition-B}).
From (\ref{CX:2}), we have  \(\sum_{x\in V^{X\times Y}_{n,y^\infty}}2^{-|x|}\leq 1\).
Let \(V'_{y^\infty}\subset\{ (x,y,k) | (x,y,k)\in V_n, y\sqsubset y^\infty\}\) such that (i) \((x,y,k),(x,y',k')\in V'_{y^\infty}\Rightarrow y=y', k=k'\) and (ii)  \((x,y,k)\in V_n, y\sqsubset y^\infty\Rightarrow \exists y'\sqsubset y^\infty, k',\ (x,y',k')\in V'_{y^\infty}\).
Then for any \(V'_{y^\infty}\) that satisfies (i) and (ii), 
from (\ref{def-Vn}) and  (\ref{condition-A0}), we have 
\[\sum_{(x,y,k)\in V'_{y^\infty}} A(x,y,k)\leq 2P(\tilde{U}^n_{y^\infty} | y^\infty) + \sum_{x\in V^{X\times Y}_{n,y^\infty}}2^{-n-|x|}<3\cdot 2^{-n}.\]
Thus  \((x,y,k)\in V_n, y\sqsubset y^\infty\Rightarrow\exists y'\sqsubset y^\infty, k', (x,y',k')\in W_n\) and hence \(V^{X\times Y}_{n,y^\infty}\subset W^{X\times Y}_{n,y^\infty}\).
From (\ref{C:3}) and (\ref{condition-A}), we have (\ref{condition-E}).
\qed

\section{Likelihood ratio test}\label{sec-ratio-test}
Let \(P\) and \(Q\) be computable probabilities on \(\Omega\).
Let 
\[r(x):=\left\{
\begin{array}{rl}
\frac{Q(x)}{P(x)},& \ \mbox{if }P(x)>0\\
0,& \ \mbox{if }P(x)=0
\end{array}\right.,\]
for \(x\in S\). 
We see that \(r\) is a computable martingale.
By the martingale convergence theorem for algorithmically random sequences \cite{takahashiIandC}, we have
\begin{col}\label{col-liklihood-conv}
\(\R^P\subset \{x^\infty\vert \lim_{x\to x^\infty}r(x)<\infty\}\).
\end{col}
The following lemma was appeared in  \cite{BM2007}.
\begin{lemma}\label{lem-classify}
Let \(P\) and \(Q\) be computable probabilities on \(\Omega\).\\
a) : \(\R^P\cap\R^Q=\R^P\cap\{x^\infty\vert 0<\lim_{x\to x^\infty}r(x)<\infty\}\).\\
b) : \(\R^P\cap (\R^Q)^c=\R^P\cap\{x^\infty\vert \lim_{x\to x^\infty}r(x)=0\}\).
\end{lemma}
Proof)
a) If \( x^\infty\in\R^P\cap\R^Q\) then \(P( x)>0\) and \(Q( x)>0\) for \( x\sqsubset x^\infty\).
From Corollary~\ref{col-liklihood-conv}, we have \(0<\lim_{ x\to x^\infty}r( x)<\infty\).
Conversely, if \( x^\infty\in\R^P\cap\{ x^\infty\vert 0<\lim_{ x\to x^\infty}r( x)<\infty\}\), by Theorem~\ref{th-levin-schnorr-multi}, \\
\(\sup_{ x\sqsubset x^\infty} -\log P( x)-Km( x)<\infty\) and \(\sup_{  x\sqsubset x^\infty} \vert -\log Q( x)+\log P( x)\vert <\infty\).
Thus,
\(\sup_{ x\sqsubset x^\infty} -\log Q( x)-Km( x)<\infty\) and we have \( x^\infty\in\R^Q\).\\
b) From a, we have \(\R^P\cap (\R^Q)^c=\R^P\cap (\R^P\cap\R^Q)^c=\R^P\cap (\{\lim r=0\}\cup\{\lim r=\infty\}\})=\R^P\cap\{\lim r=0\}\), where the last equality follows from
Corollary~\ref{col-liklihood-conv}. \qed

\begin{remark}\label{rem-ratio-0}
Let  \(g\) be an unbounded increasing total-computable function and \(\A_{\edit{g},n}:=\{(x,y) \mid |x|=n,\ (x,y)\in\A_{\edit{g}}\}\), where \(A_g\) is defined in (\ref{def:Ag}).
Let \(\F_n:=\sigma\{\Delta(x,y)\mid (x,y)\in\A_{\edit{g},n}\}\) and  \(r_n(x^\infty,y^\infty):=\frac{Q(x,y)}{P(x,y)}, (x,y)\in\A_{\edit{g}, n}\).
Then \(\{r_n\}\)  is martingale with respect to \(\{\F_n\}\).
If we replace \(\lim_{x\to x^\infty}r(x)\) with 
\(\lim_{(x,y)\to (x^\infty,y^\infty), (x,y)\in\A_{\edit{g}}(x^\infty,y^\infty)} r(x,y)\) in Corollary~\ref{col-liklihood-conv} and Lemma~\ref{lem-classify},
they  hold for computable probabilities on \(\Omega^2\).

\end{remark}

\edit{
\begin{remark}\label{rem-ratio-1}
In a similar manner with the proof of Lemma~\ref{lem-classify} a),  we have
\(\R^P\cap\R^Q=\R^P\cap\{x^\infty\vert 0<\inf_{x\sqsubset x^\infty} r(x)\}\).
If we replace \(\inf_{x\sqsubset x^\infty}\) with \(\inf_{(x,y)\in\A_g(x^\infty,y^\infty)}\) for unbounded increasing total-computable \(g\), it holds for  computable probabilities on \(\Omega^2\).
\end{remark}
}

\subsection{Absolute continuity and mutual  singularity}
By Lebesgue decomposition theorem, there exists \(N\in\edit{\B}\) such that \(P(N)=0\) and
\begin{equation}\label{eq-decom}
\forall C\in\edit{\B},\ Q(C)=\int_Cr( x^\infty)dP+Q(C\cap N).
\end{equation}
We write (a) \(P\perp Q\) if \(P\) and \(Q\) are mutually singular, i.e., there exist \(A\) and \(B\) such that 
\(A\cap B=\emptyset,\ P(A)=1\), and \(Q(B)=1\), and (b) \(P\ll Q\) if \(P\) is absolutely continuous with respect to \(Q\), i.e., 
\(\forall C\in\edit{\B}\ Q(C)=0\Rightarrow P(C)=0\).
\begin{remark}\label{rem-abs-mut}
By (\ref{eq-decom}), we have (a) \(P\perp Q\) iff \(P(\{\lim r=0\})=1\), and (b) \(P\ll Q\) iff \(P(\{\lim r=0\})=0\); for example, see \cite{neveu75}.
\end{remark}
The following theorem appeared in pp.~103 of \cite{martin-lof68} without proof.
\begin{theorem}[Martin-L\"of]\label{th-mutual-singular}
Let \(P\) and \(Q\) be computable probabilities on \(\Omega\).
Then,
\(\R^P\cap\R^Q=\emptyset\text{ iff }P\perp Q\).
\end{theorem}
Proof)
Since \(P(\R^P)=Q(\R^Q)=1\), only if part follows.
Conversely, assume that \(P\perp Q\).
Let \(N:=\{ x^\infty\vert 0<\liminf_{ x\sqsubset x^\infty}r( x)\leq\limsup_{ x\sqsubset x^\infty}r( x)<\infty\}\).
By Remark~\ref{rem-abs-mut}, we have \(P(N)=Q(N)=0\).
Since \(0<\liminf_{ x\sqsubset x^\infty}r( x)\Leftrightarrow 0<\inf_{ x\sqsubset x^\infty}r( x)\) and \(\limsup_{ x\sqsubset x^\infty}r( x)<\infty\Leftrightarrow \sup_{ x\sqsubset x^\infty}r( x)<\infty\), we have
\begin{align*}
N &=  \{ x^\infty\vert 0<\inf_{ x\sqsubset x^\infty}r( x)\leq\sup_{ x\sqsubset x^\infty}r( x)<\infty\}\\
& = \cup_{a,b\in \Qb,0<a<b<\infty}\cap_{i=1}^\infty  \tilde{N}_i^{a,b},
\end{align*}
where \(N_i^{a,b}= \{ x\vert a\leq r( y)\leq b, \forall  y\sqsubseteq x, |x|=i\}\).
Since \(P(N)=0\), we have \(\lim_i P(\tilde{N}_i^{a,b})=0\).
Since \((N_i^{a,b})^c\cap \{ x\vert P( x)>0\}\) is a r.e.~set, we can approximate \(P(\tilde{N}_i^{a,b})\) from above, and there is a computable function \(\alpha(n)\) such that 
\(P(\tilde{N}_{\alpha(n)}^{a,b})<2^{-n}\). 
Thus, \(\tilde{N}_{\alpha(n)}^{a,b}\) is a test of \(P\), and hence, \(N\subset (\R^P)^c\).
From Lemma~\ref{lem-classify} a, we have \(\R^P\cap\R^Q =\emptyset\).\qed

From Lemma~\ref{lem-classify} b and Remark~\ref{rem-abs-mut}, we have
\begin{lemma}\label{lem-abs-conti}
\(\R^P\subset\R^Q\Rightarrow P\ll Q\) for computable probabilities \(P\) and \(Q\) on \(\Omega\).
\end{lemma}
There is a counter example for the converse implication of the above lemma, see \cite{BM2007}.
The above results are related to Kakutani's theorem on product martingale \cite{{kakutani48},{williams91}}, see \cite{{fujiwara08},{vovk87b}}.

\subsection{Countable model class}
In the following discussion, let \(\{P_n\}_{n\in \Nb}\) be a family of computable probabilities on \(\Omega\); 
more precisely, we assume that there is a computable function \(A:\Nb\times S\times \Nb\to \Qb\) such that \(\vert A(n,x,k)-P_n(x)\vert< 1/k\) for all \(n,k\in \Nb\) and \(x\in S\).
Note that we cannot set \(\{P_n\}_{n\in \Nb}\) as the entire family of computable probabilities on \(\Omega\) since it is not a r.e.~set. 
Let \(\alpha\) be a computable positive probability on \(\Nb\), i.e., \(\forall n\,\alpha(n)>0\) and \(\sum_n\alpha(n)=1\).
Then, set \(P:=\sum_n\alpha(n)P_n\). We see that \(P\) is a computable probability. 
The following lemma is a special case (discrete version) of Corollary~\ref{col-sectionB}
\begin{lemma}\label{lem-disc-mix}
\(\R^P=\cup_n\R^{P_n}\).
\end{lemma}
Proof)
Let  \(P'_Y(y^\infty):=\alpha(n)\) and \(P'(x; y^\infty):=P_n(x)\)  if \(y^\infty=0^n10^\infty\) and 0 otherwise, respectively.
Let \(P'(x,y):=\int_{\Delta(y)} P'(x;y^\infty)dP'_Y\) for \(x,y\in S\), then \(P'\) is a computable probability on \(X\times Y=\Omega^2\). We see that
 \(P'_X(x)=\sum_n\alpha(n)P_n\), \(\R^{P'_Y}=\{ 0^n10^\infty | n\in\Nb\}\),   and \(\R^{P'}_{y^\infty}=\R^{P_n}\) if \(y^\infty=0^n10^\infty\).
Since \(\R^{P'_X}=\cup_{y^\infty\in\R^{P'_Y}}\R^{P'}_{y^\infty}\) (Corollary~\ref{col-sectionB}), we have the lemma.
\qed

Let \(\beta\) be a  computable probability on \(\Nb\) such that  1)  \(\beta(n)>0\) if \(n\ne n^*\) and \(\beta(n^*)=0\), and 2) \(\sum_n\beta(n)=1\).
Then, set 
\[
P^-:=\sum_n\beta(n)P_n.
\]
We see that \(P^-\) is a computable probability. 
By Lemma~\ref{lem-classify} and \ref{lem-disc-mix}, we have
\begin{col}\mbox{}\\
\(\R^{P_{n^*}}\cap_{n\ne n^*}(\R^{P_n})^c=(\cup_n \R^{P_n})\cap\{x^\infty\vert\lim_{x\to x^\infty}P^-(x)/P_{n^*}(x)=0\}\).
\end{col}
Let
\[\hat{n}(x):=\argmax_n\alpha(n)P_n(x).\]
In \cite{{barron-et-al},{barronphd}}, it is shown that 
\(\lim_{x\to x^\infty}P^-(x)/P_{n^*}(x)=0\Rightarrow \lim_{x\to x^\infty}\hat{n}(x)=n^*\).
Thus we have
\begin{col}
\(\R^{P_{n^*}}\cap_{n\ne n^*}(\R^{P_n})^c\subset \{x^\infty\vert\lim_{x\to x^\infty}\hat{n}(x)=n^*\}.\)
\end{col}
The above corollary shows that if \(x^\infty\) is random with respect to \(\R^{P_{n^*}}\) and it is not random with respect to other models then \(\hat{n}\) classifies its model. 
Estimation of models by \(\hat{n}\) is called MDL model selection, for more details, see \cite{{barron-et-al},{barronphd}}.
Note that 
by Theorem~\ref{th-mutual-singular}, if \(\{P_n\}\) are mutually singular, then \(\R^{P_{n^*}}\cap_{n\ne n^*}(\R^{P_n})^c=\R^{P_{n^*}}\), and 
by Lemma~\ref{lem-abs-conti},  if \(P_{n^*}\not\ll P^-\), then  \(\R^{P_{n^*}}\cap_{n\ne n^*}(\R^{P_n})^c\ne\emptyset\).

\section{Decomposition of complexity}\label{sec-decomp-complexity}
It can be shown that 
 \begin{equation}\label{eq-unbounded-decomp}
 \sup_{x,y\in S}~\vert Km(x,y)-Km(x\vert y)-Km(y)\vert =\infty.
 \end{equation}
The above equation shows that there is a sequence of strings such that the left-hand side of the above equation is unbounded. 
However, if we restrict strings to an increasing sequence of prefixes of random sequences \(x^\infty,y^\infty\) with respect to some computable probability and a convergence rate of conditional probability is effective,
then we can show that the left-hand-side of  (\ref{eq-unbounded-decomp}) is bounded (see Theorem~\ref{th-decompositon} below).

Let \(P\) be a computable probability on \(X\times Y=\Omega^2\). 
 From Theorem~\ref{th-ex-cond},
\begin{equation}\label{cond-conv}
\forall x, P(x\vert y)\to P(x\vert y^\infty)\mbox { as }y\to y^\infty\in\R^{P_Y}.
\end{equation}
Observe that  
\begin{equation}\label{trivial-lemma}
P(x,y)>0,\ P(x|y^\infty)>0\mbox{ if }(x,y)\sqsubset (x^\infty,y^\infty)\in\R^P.
\end{equation}
This follows from that \(P(x,y)=0\Rightarrow (x,y)\sqsubset (x^\infty,y^\infty)\notin\R^P\).
If \(P(x|y^\infty)=0\) then from (\ref{cond-conv}), we have \(\forall n\exists y\sqsubset y^\infty, P(x|y)<2^{-n}.\)
Since  \(U_n:=\{(x,y) | P(x,y)<2^{-n}P_Y(y)\}\) is a test of \(P\), we have \((x^\infty,y^\infty)\in\cap_n \tilde{U}_n\) if  \(P(x|y^\infty)=0\).

If \((x^\infty,y^\infty)\in\R^P\) then from (\ref{cond-conv}) and (\ref{trivial-lemma}), 
we have
\[
\edit{\forall x\sqsubset x^\infty, f>0\exists N\forall y\sqsubset y^\infty,\  N\leq |y|\Rightarrow  |\frac{P(x|y)}{P(x|y^\infty)}-1|<f.}
\]
By letting \(f\) be a function of \(|x|\), we have
for any \(f:\Nb\to\{q\in\Qb| q>0\}\), there is  \(g:\Nb\to\Nb\edit{\cup\{0\}}\) such that 
\begin{equation}\label{convergence-rate}
\forall (x, y)\sqsubset (x^\infty, y^\infty),\ g(|x|)\edit{\leq} |y|\Rightarrow |\frac{P(x\vert y)}{P(x\vert y^\infty)} -1 |< f(|x|).
\end{equation}
In the above, \(g\) depends on \(f\) and  \((x^\infty, y^\infty)\).
\edit{
We say that the conditional probability \(P(\cdot |y^\infty)\) is \(f, (x^\infty,y^\infty)\) {\it effectively converges } if there is a total-computable monotonically increasing \(g\) in (\ref{convergence-rate}),
where we allow that \(g\) is bounded, see Remark~\ref{rem-g}.
\(g\) is called effective convergence rate function.
}

\begin{lemma}\label{lem-cond-coding}
Let \(P\) be a computable probability on \(X\times Y=\Omega^2\) and \((x^\infty,y^\infty)\in\R^P\).
Let \(f:\Nb\to \{ q\in\Qb | 0<q<1\}\)  such that \(\sum_n f(n)<\infty\).
\edit{Assume that \(P(\cdot |y^\infty)\) is \(f, (x^\infty,y^\infty)\) effectively converges. Let \(g\) be an effective convergence rate function.}
Then there is a computable monotone function \(e: (S\cup\Omega)^2\to S\cup\Omega\) such that 
\begin{equation}\label{cond-code}
\begin{aligned}
\exists c\exists p^\infty\in\Omega &\forall (x,y)\sqsubset (x^\infty, y^\infty)\exists p\sqsubset p^\infty,\\
& g(|x|)= |y|\Rightarrow x\sqsubseteq e(p,y), |p|\leq -\log P(x | y) +c.
\end{aligned}
\end{equation}
\begin{equation}\label{cr-0}
\exists c\forall (x,y)\sqsubset (x^\infty, y^\infty),\ g(|x|)= |y|\Rightarrow Km(x | y)\leq -\log P(x | y) +c.
\end{equation}
\end{lemma}
Proof)
Let 
\begin{equation}
\begin{gathered}\label{lem-cond:A}
P'(0 | y^\infty):=P(0 | y)  \mbox{ for }|y|=g(1), y\sqsubset y^\infty,\\
P'(1 | y^\infty):=1-P'(0 |y^\infty),
\end{gathered}
\end{equation}
and for \(x\in S\) 
\begin{equation}\label{lem-cond:B}
\begin{gathered}
P'(x0 |y^\infty):=P'(x| y^\infty)\frac{P(x0 | y)}{P(x |y)}  \mbox{ if } P(x|y)>0\mbox{ for }|y|=g(|x|+1),y\sqsubset y^\infty,\\
P'(x1 |y^\infty):=P'(x| y^\infty)-P'(x0 |y^\infty).
\end{gathered}
\end{equation}

Since  \(P(x|y)>0\Leftrightarrow \forall (x',y')\sqsubseteq (x,y), P(x'|y')>0\) and \(g\) is computable,
we see that there is a partial computable  \(A:S\times S\times \Nb\to\Qb\) such that
\begin{equation}\label{cr-1}
\forall y^\infty\forall x,y,k,  |P'(x|y^\infty)-A(x,y,k)|\leq\frac{1}{k}\mbox{ if }y\sqsubset y^\infty, g(|x|)= |y|,P(x|y)>0.
\end{equation}

Let \(D:=\{(x,y)| g(|x|)= |y|,P(x|y)>0\}\).
From (\ref{cr-1}), we can construct a family of half-open intervals \(V_{(x,y)}\subset [0,1], (x,y)\in D\)  such that 
1) the end-points of \(V_{(x,y)}\) are computable with arbitrary precision form \((x,y)\in D\) and \(|V_{(x,y)}|=P'(x|y^\infty)\), and
2) if \((x,y),(x',y')\in D\), and \(y\) and \(y'\) are comparable, then  
i) \(x\sqsubseteq x'\Rightarrow V_{(x',y')}\subset V_{(x,y)}\), and 
ii) \(\Delta(x)\cap\Delta(x')=\emptyset\Rightarrow V_{(x,y)}\cap V_{(x',y')}=\emptyset\).
Let \(F:=\{(s,y,x) | I_s\subset V_{(x,y)}, (x,y)\in D\}\cup\{(s,y,\lambda) | s,y\in S\}\).  Then  \(F\) is r.e. and satisfies (\ref{monotone-condition}). 
Let \(e\) be the monotone function defined by \(F\). 
Then 
\begin{equation}\label{cr-2}
\forall y^\infty\exists c\forall x,y,\ Km_e(x | y)\leq -\log P'(x | y^\infty) +c\mbox{ if }y\sqsubset y^\infty, g(|x|)= |y|,P(x|y)>0.
\end{equation}
\edit{
By replacing  \(P(x|y)\) in (\ref{lem-cond:A}) with \(P(x|y^\infty)\), from  (\ref{convergence-rate}), we have for \(|x|=1\),
\[
(1-f(1))\leq \frac{P'(x|y^\infty)}{P(x|y^\infty)}\leq (1+f(1)).
\]
Similarly, by replacing \(P(xz| y)\) and \(P(x|y)\) in (\ref{lem-cond:B}) with \(P(xz| y^\infty)\) and \(P(x|y^\infty)\) respectively, from  (\ref{convergence-rate}), we have for \(1\leq |x|, |z|=1\),
\[
\frac{1-f(|x|+1)}{1+f(|x|)}\frac{P'(x|y^\infty)}{P(x|y^\infty)}\leq\frac{P'(xz|y^\infty)}{P(xz|y^\infty)}\leq\frac{P'(x|y^\infty)}{P(x|y^\infty)}\frac{1+f(|x|+1)}{1-f(|x|)}.
\]
Therefore we have
\[
\frac{\prod_{n=1}^{|x|} (1-f(n))}{\prod_{n=1}^{|x|-1}(1+f(n))}\leq \frac{P'(x|y^\infty)}{P(x|y^\infty)}\leq \frac{\prod_{n=1}^{|x|} (1+f(n))}{\prod_{n=1}^{|x|-1}(1-f(n))}\text{ if }P(x|y^\infty)>0.
\]
Since \(0<\prod_{n=1}^{\infty}(1-f(n))\leq \prod_{n=1}^{\infty}(1+f(n))<\infty\) if \(\sum_n f(n)<\infty\) and \(0<f<1\), from (\ref{cr-2}), we have the lemma.}
\qed

\begin{theorem}\label{th-decompositon}
Let \(P\) be a computable probability on \(X\times Y=\Omega^2\) and \((x^\infty,y^\infty)\in\R^P\).
Let \(f:\Nb\to \{ q\in\Qb | 0<q<1\}\)  such that \(\sum_n f(n)<\infty\).
\edit{Assume that \(P(\cdot |y^\infty)\) is \(f, (x^\infty,y^\infty)\) effectively converges. Let \(g\) be an effective convergence rate function.}
Then
\begin{equation}\label{deco-1}
\sup_{(x,y)\in \A_{\edit{g}}(x^\infty, y^\infty)} | Km(x | y)+\log P(x | y) |<\infty,
\end{equation}
\begin{equation}\label{deco-3}
\sup_{(x,y)\in \A_{\edit{g}}(x^\infty, y^\infty)}  |Km(x,y)-Km(x | y)-Km(y)|<\infty,
\end{equation}
\edit{where \(A_g\) is defined in (\ref{def:Ag}).}
In addition, if \(P(\cdot |y^\infty)\) is computable relative to \(y^\infty\), then 
\begin{equation}\label{deco-2}
\sup_{(x,y)\in \A_{\edit{g}}(x^\infty, y^\infty)}  Km(x | y)-Km(x|y^\infty )<\infty.
\end{equation}
\end{theorem}
Proof)
From Corollary~\ref{col-coding}, if \((x^\infty,y^\infty)\in\R^P\) then
\begin{equation}\label{sub-deco}
\sup_{(x,y)\in\A_{\edit{g}}(x^\infty,y^\infty)} |\log P(x,y)+Km(x,y)|<\infty, \sup_{y\sqsubset y^\infty}|\log P_Y(y)+Km(y)|<\infty,
\end{equation}
\begin{equation}\label{deco-lower}
\sup_{(x,y)\in\A_{\edit{g}}(x^\infty,y^\infty)} -\log P(x|y)-Km(x|y)<\infty.
\end{equation}
From  (\ref{cr-0}) and (\ref{deco-lower}), we have (\ref{deco-1}).
From (\ref{sub-deco}) and (\ref{deco-1}), we have (\ref{deco-3}).

If \(P(\cdot |y^\infty)\) is computable relative to \(y^\infty\), Theorem~\ref{th-general-Lambalgen} holds, i.e., \((x^\infty,y^\infty)\in\R^P\mbox{ iff } x^\infty\in\R^{P(\cdot |y^\infty),y^\infty}, y^\infty\in\R^{P_Y}\).
By relativized version of Levin-Schnorr theorem, we have
\[\sup_{x\sqsubset x^\infty} -\log P(x|y^\infty)-Km(x|y^\infty)<\infty,\]
for \((x^\infty,y^\infty)\in\R^P\).
Since \(Km(x|y^\infty)\leq Km(x|y)\) for \(y\sqsubset y^\infty\), from (\ref{convergence-rate}) and  (\ref{deco-1}), we have (\ref{deco-2}).
\qed
\begin{example}\label{exampleA}
Let \(P'\) be a computable probability on \(\Omega\).
For \(x=x_1\cdots x_n,y=y_1\cdots y_m\in S\), let 
\[
\begin{array}{ll}
\Delta(x\oplus y):=\{z_1z_2\cdots\in\Omega\ | & z_i=x_i\mbox{ if }i\mbox{ is odd and }i\leq n,\\
                                                                                & z_i=y_i\mbox{ if }i\mbox{ is even and }i\leq m\},
\end{array}
\]
\[P(x,y):=P'(\Delta(x\oplus y)).\]
Then \(P\) is a computable probability on \(X\times Y=\Omega^2\), i.e., \(X\) and \(Y\) are the spaces of odd and even coordinates, respectively. 
For \(x^\infty=x_1x_2\cdots, y^\infty=y_1y_2\cdots\),
let 
\[x^\infty\oplus y^\infty:=x_1y_1x_2y_2\cdots\in\Omega,\]
then
\[(x^\infty,y^\infty)\in\R^P\iff x^\infty\oplus y^\infty\in\R^{P'}.\]
From Theorem~\ref{th-general-Lambalgen}, if the conditional probability is computable relative to \(y^\infty\in\R^{P_Y}\) then
\(x^\infty\oplus y^\infty\) is random with respect to \(P'\) iff \(y^\infty\) is random and \(x^\infty\) is random with respect to the conditional probability at \(y^\infty\).
Let \(P'\) be a computable first order Markov process, i.e., \(P'(z_1\cdots z_n)=p_{z_1}\Pi_{i=2}^n p_{z_{i-1},z_i}\),
where \(\forall i,j\in\{0,1\}, 0\leq p_i, p_{i,j}\leq 1, \sum_i p_i=1, \sum_j p_{i,j}=1\).
We see that  \(P(x|y^\infty)=P(x| y_1\cdots y_{|x|})\). Thus  \(g(n)=n\) satisfies  (\ref{convergence-rate}) for any \(f\) and Theorem~\ref{th-decompositon} holds.
\end{example}
\edit{
\begin{remark}\label{rem-g}
In Lemma~\ref{lem-cond-coding} and Theorem~\ref{th-decompositon}, \(g\) need not be unbounded if (\ref{convergence-rate}) hold.
For example if  \(P:=P_XP_Y\) then \(g:=0\) satisfies (\ref{convergence-rate}) for any \(f\).
\end{remark}
}

\subsection{Independence}
We show some equivalent conditions for independence of two individual sequences.
\edit{
The following result shows that if \((x^\infty,y^\infty)\) is random with respect to some computable probability (in \cite{muchnik98} such a sequence is called {\it natural}), then we can represent  independence of \((x^\infty,y^\infty)\) in terms of complexity.
}

\begin{col}\label{col-independence}
Let \(P\) be a computable probability on \(\Omega^2\) and  \((x^\infty,y^\infty)\in\R^P\).
Assume that \(P(\cdot |y^\infty)\) is computable relative to \(y^\infty\) and 
\edit{\(f, (x^\infty,y^\infty)\) effectively converges for \(f=1\).}
Let \(Q\) be a computable probability such that \(\forall x,y, Q(x,y):=P_X(x)P_Y(y)\).
The following statements  are equivalent:\\
a) \((x^\infty,y^\infty)\in\R^Q\).\\
b) 
for any unbounded computable increasing \(g\),\\
\(\sup_{(x,y)\in\A_{\edit{g}}(x^\infty,y^\infty)}\vert Km(x,y)-Km(x)-Km(y)\vert<\infty.\)\\
c) 
\(\sup_{x\sqsubset x^\infty}Km(x)-Km(x | y^\infty)<\infty.\)
\end{col}
Proof)
a\(\Rightarrow\)b:
Every increasing computable \(g\) satisfies (\ref{convergence-rate}) for \(Q\).
From Theorem~\ref{th-decompositon}, if \((x^\infty,y^\infty)\in\R^Q\) then
\(\sup_{(x,y)\in \A_{\edit{g}}(x^\infty, y^\infty)} | Km(x | y)+\log P_X(x) |<\infty\), \(\sup_{x\sqsubset x^\infty} |Km(x)+\log P_X(x)|<\infty\), and (\ref{deco-3}) holds.
Thus we have b.\\
b\(\Rightarrow\)a: \edit{Let \(g\) be unbounded computable increasing function.} Since \(\R^P\subset\R^{P_X}\times\R^{P_Y}\), 
\[
\begin{array}{ll}
(x^\infty,y^\infty)\in\R^P\Rightarrow & \sup_{(x,y)\in\A_{\edit{g}}(x^\infty,y^\infty)} | Km(x,y)+\log P(x,y)|<\infty, \\
& \sup_{x\sqsubset x^\infty}|Km(x)+\log P_X(x)|<\infty, \\
& \sup_{y\sqsubset y^\infty}|Km(y)+\log P_Y(y)|<\infty.
\end{array}
\]
We have
\edit{\(0<\inf_{ (x,y)\in\A_g(x^\infty,y^\infty)}\frac{Q(x,y)}{P(x,y)}\).
From Lemma~\ref{lem-classify} (see Remark~\ref{rem-ratio-1}), we have a.}\\
 a\(\Rightarrow\)c:
 \edit{Since \(g:=0\) satisfies (\ref{convergence-rate}) for \(Q\), from Theorem~\ref{th-decompositon}, we have c, see Remark~\ref{rem-g}.}\\
c\(\Rightarrow\)a:
\edit{Let \(g\) be an unbounded effective convergence rate function for \(P(\cdot |y^\infty), f=1,\text{ and }(x^\infty,y^\infty)\in\R^P\).}
Then we have \(\frac{P_X(x)}{2P(x|y^\infty)}\leq \frac{P_X(x)}{P(x|y)}=\frac{Q(x,y)}{P(x,y)}\) for \((x,y)\in \A_{\edit{g}}(x^\infty,y^\infty)\).
From Theorem~\ref{th-general-Lambalgen} and Levin-Schnorr theorem, we have\\
\(\sup_{x\sqsubset x^\infty}|Km(x|y^\infty)+\log P(x|y^\infty)|<\infty\) and \(\sup_{x\sqsubset x^\infty}|Km(x)+\log P_X(x)|<\infty\).
From the statement c), we have \(0<\edit{\inf_{(x,y)\in\A_{g}(x^\infty,y^\infty)}\frac{Q(x,y)}{P(x,y)}}\).
From Lemma~\ref{lem-classify} (see Remark~\ref{rem-ratio-1}), we have a.
\qed

\edit{Note that  \(\R^P\cap\R^Q\ne\emptyset\) iff \(P\) and \(Q\) are not mutually singular (Theorem~\ref{th-mutual-singular}) iff \(P(\lim r>0)>0\) (Remark~\ref{rem-abs-mut}).}

\section{Bayesian statistics}\label{sec-bayes}
Let \(P\) be a computable probability on \(X\times Y\) and \(P_X,~P_Y\) be its marginal distributions as before. 
In Bayesian statistical terminology, if \(X\) is a sample space, then \(P_X\) is called mixture distribution, and if \(Y\) is a parameter space, then \(P_Y\) is called prior distribution.
We show that section of random set satisfies many theorem of Bayesian statistics, see also \cite{takahashiIandC}, and it is natural as a definition of random set with respect to conditional probability from
Bayesian statistical point of view.

\subsection{Consistency of posterior distribution}
\begin{figure}[h]
\setlength{\unitlength}{0.8mm}
\begin{picture}(80,70)(-20,0)
\put(35,10){\framebox(60,60)}
\put(65,0){\(\R^{P_Y}\)}
\put(20,47){\(\R^{P_X}\)}
\put(95,0){\(Y\)}
\put(25,70){\(X\)}
\put(48,25){\(\R^P\)}
\put(65,10){\line(0,1){60}}
\put(35,60){\line(1,0){60}}
\put(75,10){\circle*{1}}
\put(35,60){\circle*{1}}
\put(75,60){\circle*{1}}
\put(75,40){\circle*{1}}
\put(75,10){\line(0,1){60}}
\put(35,50){\line(1,0){60}}
\put(35,40){\line(1,0){60}}
\put(65,10){\circle*{1}}
\put(35,50){\circle*{1}}
\put(65,50){\circle*{1}}
\put(35,40){\circle*{1}}
\end{picture}
\caption{\(f:\R^{P_X}\to\R^{P_Y}\)}
\label{fig-1}
\end{figure}
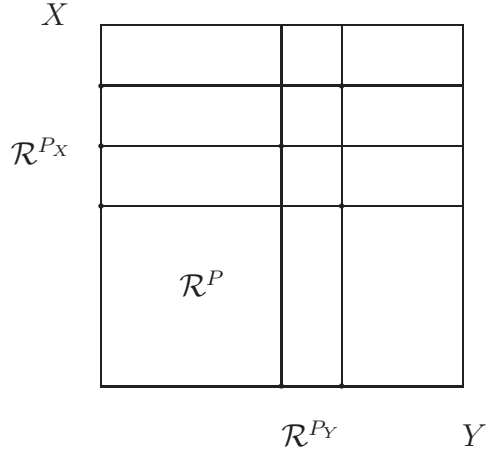

We show a consistency of posterior distribution for algorithmically random sequences. 
We see that the classification of random sets by likelihood ratio test  (see Section~\ref{sec-ratio-test}) plays an important role in this section. 
\begin{theorem}\label{th-consis-pos}
Let \(P\) be a computable probability on \(X\times Y=\Omega^2\).
The following six statements are equivalent:\\
a) \( P(\cdot\vert  y)\perp  P(\cdot\vert  z)\) if \(\Delta( y)\cap\Delta( z)=\emptyset, P_Y(y)>0, P_Y(z)>0\) for \( y, z\in S\).\\
b) \(\R^{ P(\cdot\vert  y)}\cap\R^{ P(\cdot\vert  z)}=\emptyset\) if \(\Delta( y)\cap\Delta( z)=\emptyset, P_Y(y)>0, P_Y(z)>0\)  for \( y, z\in S\).\\
c)  \( P_{Y|X}(\cdot\vert x)\) converges weakly to \(I_{ y^\infty}\) as \(x\to x^\infty\) for \((x^\infty, y^\infty)\in\R^{ P}\), where \(I_{ y^\infty}\) is the distribution that has probability of 1 at \( y^\infty\).\\
d) \(\R^{ P}_{ y^\infty}\cap\R^{ P}_ {z^\infty}=\emptyset\) if \( y^\infty\ne z^\infty\).\\
e) There exists a surjective function \(f:\R^{P_X}\to \R^{P_Y}\) such that  \(f(x^\infty)= y^\infty\) for \((x^\infty, y^\infty)\in\R^{ P}\).\\
f) There exists \(f:X\to Y\) and \( Y'\subset Y\) such that \(P_Y( Y')=1\) and \(f= y^\infty,\ P(\cdot|  y^\infty)-a.s.\) for  \( y^\infty\in Y'\).\\
\end{theorem}
Proof)
a \(\Leftrightarrow\) b follows from Theorem~\ref{th-mutual-singular}.\\
b \(\Rightarrow\) c :  
If \((x^\infty, y^\infty)\in\R^{ P}\), then \(x^\infty\in\R^{ P(\cdot\vert  y)}\) and \(P_Y(y)>0\) for \(y\sqsubset y^\infty\).
If \(\Delta( y)\cap\Delta( z)=\emptyset\) and \(P_Y(z)>0\), then  from the statement b, \(x^\infty\notin\R^{ P(\cdot\vert  z)}\).
If \(P_Y(z)>0\) then from Lemma~\ref{lem-classify}, we have \(\lim_{x\to x^\infty} P(x\vert  z)/ P(x\vert  y)=0\), and
\begin{eqnarray}
\lim_{x\to x^\infty}\frac{ P(x\vert  z)}{ P(x\vert  y)}=0 \Leftrightarrow  \lim_{x\to x^\infty}\frac{ P(x, z)}{ P(x, y)}=0\Leftrightarrow  \lim_{x\to x^\infty}\frac{ P_{Y|X}( z\vert x)}{ P_{Y|X}( y\vert x)}=0.\label{sub-eq-consis}
\end{eqnarray}
If \(P_Y(z)=0\) then \edit{the last equation in }(\ref{sub-eq-consis}) holds. 
Hence \edit{the last equation in }(\ref{sub-eq-consis}) holds for all \(z\) and
we see that the posterior distribution \( P_{Y|X}(\cdot\vert x)\) converges weakly to \(I_{ y^\infty}\).\\
c  \(\Rightarrow\) d :  obvious.\\
d \(\Rightarrow\) e  :  Since \(\R^{ P}_{ y^\infty}\cap\R^{ P}_{ z^\infty}=\emptyset\) for \( y^\infty\ne z^\infty\),
we can define a function \(f:X\to Y\) such that  \(f(x^\infty)= y^\infty\) for \(x^\infty\in\R^{ P}_{ y^\infty}\).
From Corollary~\ref{col-sectionB}, we have e, see Figure~\ref{fig-1}.\\
e \(\Rightarrow\) f : By Theorem~\ref{col-sectionA}, we have f.\\
f \(\Rightarrow\) a : Let \(A_{ y^\infty}:=\{x^\infty\vert f(x^\infty)= y^\infty\}\). Then, \(A_{ y^\infty}\cap A_{ z^\infty}=\emptyset\) for \( y^\infty\ne z^\infty\) and
\(P(A_{ y^\infty}| y^\infty)=1\) for \( y^\infty\in Y'\).
Thus, 
\((\cup_{ y^\infty\in\Delta( y)}A_{ y^\infty})\cap (\cup_{ y^\infty\in\Delta( z)}A_{ y^\infty})=\emptyset\) for \(\Delta( y)\cap\Delta( z)=\emptyset\) and 
\( P(\cup_{ y^\infty\in\Delta( y)}A_{ y^\infty}\vert  y)= P(\cup_{ y^\infty\in\Delta( z)}A_{ y^\infty}\vert  z)=1\), which shows a.
\qed\\

Usually, consistency of posterior distribution is derived from f, see \cite{doob48}.
Note that the statements a and f do not contain algorithmic notion. 
\begin{example}
Let \(\{P(\cdot;  y^\infty)\}_{y^\infty\in Y}\) be the parametric model of Bernoulli process, i.e.,
\(P(x; y^\infty):=r(y^\infty)^{\sum_{i=1}^n x_i}(1-r(y^\infty))^{n-\sum_{i=1}^n x_i}\) where \(x=x_1\cdots x_n\), \(y^\infty=y_1y_2\cdots\), and \(r(y^\infty):=\sum_i y_i2^{-i}\).
Let \(P_Y\) be a computable probability on \(\Omega\) and \(P(x,y):=\int_{\Delta(y)}P(x;y^\infty)dP_Y\) for \(x,y\in S\).
Then \(P\) is a computable probability on \(\Omega^2\).
By the law of large numbers, f (and all the statements) are satisfied.
Note that the conditional probability \(P(\cdot |y^\infty)\) is defined by \(P\), see Section~4 in \cite{takahashiIandC}. In general, it is possible that \(P(\cdot |y^\infty)\ne P(\cdot; y^\infty)\) at \(y^\infty\) of a null set.
\end{example}

\subsection{Algorithmically best estimator}
We study asymptotic theory of estimation for individual samples and parameters from algorithmic point of view. 

Suppose that one of the statement of Theorem~\ref{th-consis-pos} holds.
Then from the statement c, we have  \(P(y |x^\infty)=1\) for \(y\sqsubset y^\infty, (x^\infty,y^\infty)\in\R^P\).
Since \(P(y|x)\to P(y|x^\infty)\) as \(x\to x^\infty\) if \(x^\infty\in\R^{P_X}\), we have
\(\forall \epsilon>0, y\sqsubset y^\infty, \exists x\sqsubset x^\infty, P(y|x)>1-\epsilon\).
In particular there is an increasing \(h\) such that 
\(\forall \epsilon, y\sqsubset y^\infty,  x\sqsubset x^\infty, |x|\geq h(|y|)\Rightarrow P(y|x)>1-\epsilon\).
Roughly speaking, the following theorem shows that if this happen then \(y\) is estimated from \(x\) of  size \(h\) and 
if \(P(y|x)\) goes to 0 then we cannot estimate \(y\) from sample size \(h\).

\begin{theorem}\label{th-estimation}
Let \(P\) be a computable probability on \(X\times Y=\Omega^2\).
Let  \(h:\Nb\to\Nb\) be an increasing computable function and  \(\A:=\{ (x,y) | |x|=h(|y|)\}\).
For each \((x^\infty,y^\infty)\) we have: \\
a) 
If \(\inf_{(x,y)\in\A(x^\infty,y^\infty)} P(y|x)>0\), then there is a computable function \(\rho\) such that
\(y=\rho(x)\) for infinitely many \((x,y)\in\A(x^\infty,y^\infty)\), where \(\rho\) need not be monotone.\\
b) 
\edit{Let \(f:\Nb\to \{ q\in\Qb | 0<q<1\}\)  such that \(\sum_n f(n)<\infty\).}
Assume that  
\edit{\(P(\cdot | x^\infty)\) effectively converges for \(f\) and  \((x^\infty,y^\infty)\in\R^P\), i.e., 
 there is a total computable increasing \(h:\Nb\to\Nb\) such that }
\[
|x|=h(|y|)\Rightarrow |\frac{P(y\vert x)}{P(y\vert x^\infty)} -1 |<f(|y|).
\]
If  \(\inf_{(x,y)\in\A(x^\infty,y^\infty)} P(y|x)>0\) then
there is a computable monotone function \(\rho\)  such that \(\forall (x,y)\in\A(x^\infty,y^\infty), y\sqsubseteq\rho(x)\).\\
c) If \((x^\infty,y^\infty)\in\R^{ P}\) and  \(\inf_{(x,y)\in\A(x^\infty,y^\infty)} P(y|x)=0\), then
there is no computable monotone function \(\rho\) such that  \(\forall (x,y)\in\A(x^\infty,y^\infty), y\sqsubseteq\rho(x)\).
\end{theorem}
Proof) 
a) 
By applying Shannon-Fano-Elias coding to \(P(\cdot\vert x)\) on the finite partition \(\{y\vert \vert y\vert =h^{-1}(\vert x\vert)\}\),
we can construct a computable function \(e\) and a program \(p\in S\)
such that \(e(p,x)= y\) and \(\vert p\vert=\lceil -\log P(y\vert x)\rceil+1\).
Here, \(e\) need not be a monotone function. 
Since \(\vert p\vert<\infty\) as \(x\to x^\infty\), there is a \(p_0\) such that \(e(p_0, x)=y\) for infinitely many prefix \(x\) of \(x^\infty\).
Thus, \(\rho(x):=e(p_0,x)\) satisfies a.\\
b) 
From  (\ref{cond-code}), there is a computable monotone function \(e\) and \(p\in S\) such that 
\(\forall (x,y)\sqsubset \A(x^\infty,y^\infty), y\sqsubseteq e(p,x)\).
Let \(\rho(x):=e(p,x)\) then \(\rho\) satisfies b.\\
c) 
As in the same way of (\ref{deco-lower}), we have 
\(\sup_{(x,y)\in\A(x^\infty,y^\infty)} -\log P(y|x)-Km(y|x)<\infty\).
Since \(\sup_{(x,y)\in\A(x^\infty,y^\infty)} -\log P(y|x)=\infty\), we have  \\
\(\sup_{(x,y)\in\A(x^\infty,y^\infty)} Km(y|x)=\infty\).
If there is a computable monotone function \(\rho\) such that \(\forall (x,y)\in\A(x^\infty,y^\infty), y\sqsubseteq\rho(x)\) then
  \(\sup_{(x,y)\in\A(x^\infty,y^\infty)} Km(y|x)<\infty\), which is a contradiction.\qed

By definition, we have
\begin{equation}\label{eq-redund}
-\log P(y\vert x)=-\log\int_{\Delta(y)}P(x|y^\infty)dP_Y(y^\infty)+\log\int_ Y P(x|y^\infty)dP_Y(y^\infty).
\end{equation}
Let \(P_Y\) be a Lebesgue absolutely continuous measure.
Let \(\hat{y}\) be the maximum likelihood estimator.
By using Laplace approximation with suitable conditions, if \(\hat{y}\in\Delta(y)\) and \(h^{-1}(\vert x\vert)\approx \frac{1}{2}\log \vert x\vert\), then 
the right-hand-side of (\ref{eq-redund}) is asymptotically bounded, for example see \cite{barron-et-al}, and we have \(\inf_{x\sqsubset x^\infty} P(y\vert x)>0\),
where \(|y|=h^{-1}(|x|)\).
Thus, by Theorem~\ref{th-estimation} a, we can compute initial \(\lceil\frac{1}{2}\log \vert x\vert\rceil\)-bits of \(y^\infty\) from \(x\) infinitely many times, 
which is an algorithmic version of a well known result in statistics: \(\vert y^\infty-\hat{y}\vert=O(1/\sqrt{n})\).

Let \(h^{-1}(\cdot)\) be a large order function such that 
\(\inf_{x\sqsubset x^\infty} P(y\vert x)=0\) for \(|y|=h^{-1}(|x|)\); for example, set \(h^{-1}(\vert x\vert)=\lceil\log \vert x\vert\rceil\).
By Theorem~\ref{th-estimation} c, there is no monotone computable function that computes initial \(h^{-1}(\vert x\vert)\)-bits of \(y^\infty\) for all \(x\sqsubset x^\infty\).
If such a function exists, then \(y^\infty\) is not random with respect to \(P_Y\) and the Lebesgue measure of such parameters is  \(0\).
On the other hand, it is known that the set of parameters that are estimated within \(o(1/\sqrt{n})\) accuracy has Lebesgue measure \(0\) \cite{lecam53}.

Theorem~\ref{th-estimation} shows a relation between the redundancy of universal coding and parameter estimation; as in \cite{takahasiIT2004}, if we set \(P_Y\) to be a singular
prior, we have \(\inf_{x\sqsubset x^\infty} P(y\vert x)>0\) for a large order \(h^{-1}\).
In such a case we have a super-efficient estimator.

\begin{center}
{\bf Acknowledgement}
\end{center}
The author thanks Prof.~Teturo Kamae (Matsuyama Univ.), Prof.~Akio Fujiwara,  and Masahiro~Nakamura (Osaka Univ.) for discussions and comments to earlier versions of the paper. 
The author also thanks anonymous referees for valuable comments, which helped to improve the paper. 
{\small
\bibliographystyle{plain}

}
\end{document}